\documentclass[12pt]{article}
\usepackage{amsfonts}
\usepackage{amssymb}
\usepackage{graphics,psboxit,amsmath}
\usepackage{subfigure}
\usepackage{graphicx}
\usepackage{verbatim}
\usepackage{epic}

\def\hybrid{\topmargin 0pt \oddsidemargin 0pt 
        \headheight 0pt \headsep 0pt
        \textwidth 16,5cm 
        \textheight 23cm 
        \marginparwidth .875in
        \parskip 5pt plus 1pt \jot = 1.5ex}

\hybrid

\def\baselinestretch{1.2}

\catcode`\@=11

\def\marginnote#1{}
\newcount\hour
\newcount\minute
\newtoks\amorpm
\hour=\time\divide\hour by60
\minute=\time{\multiply\hour by60 \global\advance\minute by-\hour}
\edef\standardtime{{\ifnum\hour<12 \global\amorpm={am}%
        \else\global\amorpm={pm}\advance\hour by-12 \fi
        \ifnum\hour=0 \hour=12 \fi
        \number\hour:\ifnum\minute<10 0\fi\number\minute\the\amorpm}}
\edef\militarytime{\number\hour:\ifnum\minute<10 0\fi\number\minute}

\def\draftlabel#1{{\@bsphack\if@filesw {\let\thepage\relax
   \xdef\@gtempa{\write\@auxout{\string
      \newlabel{#1}{{\@currentlabel}{\thepage}}}}}\@gtempa
   \if@nobreak \ifvmode\nobreak\fi\fi\fi\@esphack}
        \gdef\@eqnlabel{#1}}
\def\@eqnlabel{}
\def\@vacuum{}
\def\draftmarginnote#1{\marginpar{\raggedright\scriptsize\tt#1}}

\def\draft{\oddsidemargin -.5truein
        \def\@oddfoot{\sl preliminary draft \hfil
        \rm\thepage\hfil\sl\today\quad\militarytime}
        \let\@evenfoot\@oddfoot \overfullrule 3pt
        \let\label=\draftlabel
        \let\marginnote=\draftmarginnote
   \def\@eqnnum{(\theequation)\rlap{\kern\marginparsep\tt\@eqnlabel}%
\global\let\@eqnlabel\@vacuum} }

\def\draft2{
        \def\@oddfoot{\sl preliminary draft \hfil
        \rm\thepage\hfil\sl\today\quad\militarytime}
        \let\@evenfoot\@oddfoot \overfullrule 3pt
        \let\label=\draftlabel
        \let\marginnote=\draftmarginnote
   \def\@eqnnum{(\theequation)\rlap{\kern\marginparsep\tt\@eqnlabel}%
\global\let\@eqnlabel\@vacuum} }

\def\preprint{\twocolumn\sloppy\flushbottom\parindent 2em
        \leftmargini 2em\leftmarginv .5em\leftmarginvi .5em
        \oddsidemargin -.5in \evensidemargin -.5in
        \columnsep .4in \footheight 0pt
        \textwidth 10.in \topmargin -.4in
        \headheight 12pt \topskip .4in
        \textheight 6.9in \footskip 0pt
        \def\@oddhead{\thepage\hfil\addtocounter{page}{1}\thepage}
        \let\@evenhead\@oddhead \def\@oddfoot{} \def\@evenfoot{} }

\def\numberbysection{\@addtoreset{equation}{section}
        \def\theequation{\thesection.\arabic{equation}}}

\def\underline#1{\relax\ifmmode\@@underline#1\else
        $\@@underline{\hbox{#1}}$\relax\fi}

\def\titlepage{\@restonecolfalse\if@twocolumn\@restonecoltrue\onecolumn
     \else \newpage \fi \thispagestyle{empty}\c@page\z@
        \def\thefootnote{\fnsymbol{footnote}} }

\def\endtitlepage{\if@restonecol\twocolumn \else \newpage \fi
        \def\thefootnote{\arabic{footnote}}
        \setcounter{footnote}{0}} 

\catcode`@=12
\relax

\def\figcap{\section*{Figure Captions\markboth
        {FIGURECAPTIONS}{FIGURECAPTIONS}}\list
        {Figure \arabic{enumi}:\hfill}{\settowidth\labelwidth{Figure
999:}
        \leftmargin\labelwidth
        \advance\leftmargin\labelsep\usecounter{enumi}}}
 \relax
\def\tablecap{\section*{Table Captions\markboth
        {TABLECAPTIONS}{TABLECAPTIONS}}\list
        {Table \arabic{enumi}:\hfill}{\settowidth\labelwidth{Table
999:}
        \leftmargin\labelwidth
        \advance\leftmargin\labelsep\usecounter{enumi}}}
 \relax
\def\reflist{\section*{References\markboth
        {REFLIST}{REFLIST}}\list
        {[\arabic{enumi}]\hfill}{\settowidth\labelwidth{[999]}
        \leftmargin\labelwidth
        \advance\leftmargin\labelsep\usecounter{enumi}}}
 \relax
\makeatletter
\newcounter{pubctr}
\def\publist{\@ifnextchar[{\@publist}{\@@publist}}
\def\@publist[#1]{\list
        {[\arabic{pubctr}]\hfill}{\settowidth\labelwidth{[999]}
        \leftmargin\labelwidth
        \advance\leftmargin\labelsep
        \@nmbrlisttrue\def\@listctr{pubctr}
        \setcounter{pubctr}{#1}\addtocounter{pubctr}{-1}}}
\def\@@publist{\list
        {[\arabic{pubctr}]\hfill}{\settowidth\labelwidth{[999]}
        \leftmargin\labelwidth
        \advance\leftmargin\labelsep
        \@nmbrlisttrue\def\@listctr{pubctr}}}
 \relax
\makeatother

\def\ba{\begin{equation}}
\def\ea{\end{equation}}

\def\no{\noindent}

\def\IR{\relax{\rm I\kern-.18em R}}

\begin{document}


\renewcommand{\theequation}{\thesection.\arabic{equation}}
\csname @addtoreset\endcsname{equation}{section}

\newcommand{\eqn}[1]{(\ref{#1})}
\newcommand{\be}{\begin{eqnarray}}
\newcommand{\ee}{\end{eqnarray}}
\newcommand{\non}{\nonumber}
\begin{titlepage}
\strut\hfill
\vskip 1.3cm
\begin{center}


{\large \bf Defects in the discrete non-linear Schr\"{o}dinger model}

\vskip 0.5in

{\bf  Anastasia Doikou}\\
\vskip 0.3in

{\footnotesize
University of Patras, Department of Engineering Sciences, Physics Division \\
GR-26500 Patras, Greece}\\

\vskip .2in


{\footnotesize {\tt E-mail: adoikou$@$upatras.gr}}\\

\end{center}

\vskip .6in

\centerline{\bf Abstract}

The discrete non-linear Schr\"{o}dinger (NLS) model in the presence of an integrable defect is examined.
The problem is viewed from a purely algebraic point of view, starting from the fundamental algebraic relations
that rule the model. The first charges in involution are explicitly constructed, as well as the corresponding Lax pairs.
These lead to sets of difference equations, which include particular terms corresponding to the impurity point.
A first glimpse regarding the corresponding continuum limit is also provided.

\no

\vfill
\no


\end{titlepage}
\vfill
\eject


\tableofcontents

\def\baselinestretch{1.2}
\baselineskip 20 pt
\no

\section{Introduction}

The presence of defects in $1+1$ integrable field theories has been the subject of intense research during the recent years (see e.g. \cite{delmusi}--\cite{weston}). It is well established by now that the requirement of integrality leads to a set of severe algebraic constraints that should be satisfied by the associated degrees of freedom as well as the relevant physical quantities, such as scattering matrices, at the quantum level (see e.g. \cite{delmusi, fus}). In integrable field theories the defect is usually introduced as a discontinuity together with suitable sewing conditions \cite{konle}--\cite{haku}. In this case no systematic algebraic description exist so far with the exception of some recent attempts \cite{haku}, but again the issue of integrability is not fully resolved to our understanding (see also \cite{caudr}).

In the present study we start our investigation using an integrable model on the one dimensional lattice, that is the discrete non-linear Schr\"{o}dinger (NLS) model, and impose an ultra-local integrable defect (see also relevant considerations in \cite{weston}). We by construction deal with an integrable system by a priori imposing the necessary algebraic constraints that ensure the integrability of the model. We then explicitly construct the first integrals of motion as well as the relevant Lax pairs. The corresponding equations of motion are also derived.
The pertinent question in this frame is whether and how integrability is preserved in the continuum limit. In other words is the underlying algebra that ensures integrability modified and how? We make a first attempt to answer these questions by considering the continuum limits of the first couple of integrals of motion as well as the corresponding Lax pairs; then certain continuity or sewing conditions are naturally induced via the process. These preliminarily results on the continuum limit, provide a first insight on how a systematic continuum process should be formulated within this context.

The outline of the present article is as follows: In the next section we briefly review the discrete NLS model with periodic boundary conditions. In section 3 the first three integrals of motion are derived together with the corresponding Lax pairs. We also derive the equations of motion associated to the third charge, which is the typical Hamiltonian. In the next section we consider the discrete NLS model in the presence of an integrable local defect. We introduce a suitable Lax operator associated to the defect point so that the system is by construction integrable. We then derive the first integrals of motion as well as the corresponding Lax pairs. These are novel expressions that contain non-trivial contributions due to the presence of the integrable defect. The related equations of motion for the Hamiltonian are also derived. Particular emphasis is given exactly on the defect point where the equations of motion are of a completely different form compared to the other points due to the structural dissimilarity of the associated Lax operator.
Finally, in section 5 we provide a first glimpse on the continuum limits of the derived physical quantities. Certain sewing conditions naturally arise as continuity requirements, ensuring the Poisson commutativity of the first two continuum charges.

\section{The periodic DNLS model }

We shall briefly review  the discrete NLS model with periodic boundary conditions.
We shall reproduce the first three local integrals of motion and the associated Lax pairs.
In the subsequent sections we shall repeat these derivations in the presence of an integrable defect.
Our ultimate aim is to make a contact with recent results on defects arising in 1+1 integrable field theories (see e.g. \cite{cozanls}).
We aim at taking the appropriate continuum limit, that will provide the classical continuum models with defects that are still integrable. This is conceptually and technically a very intriguing problem,
and will be pursued in full detail in forthcoming investigations. Here, however we provide some preliminary results regarding the continuum limit of the first two integrals of motion.

\section{Local Integrals of motion}

Our main aim in this section is to extract the first integrals of motion for the periodic discrete NLS model. The associated Lax operator is given by (see e.g. \cite{kundura}):
\be
L_{aj}(\lambda) &=& \lambda D_j + A_j \non\\&=&  \begin{pmatrix}
   \lambda +{\mathbb N}_j  & x_j \\
    -X_j & 1
\end{pmatrix} \ \label{lmatrix}
\ee where ${\mathbb N}_j = 1- x_j X_j$.

We shall focus here, mainly for simplicity, on the classical case.
However, we have to note that our results are valid in the quantum case as well.
The $L$ matrix satisfies the fundamental algebraic relation \cite{FT}\footnote{The same $L$-operator holds for the quantum case as well. It then satisfies:
\be
R_{12}(\lambda_1 -\lambda_2)\ L_1(\lambda_1)\ L_2(\lambda_2) = L_2(\lambda_2)\ L_1(\lambda_1)\ R_{12}(\lambda_1- \lambda_2),
\ee where $R(\lambda)= \lambda {\mathbb I} + {\cal P}$.}
\be
\{L_a(\lambda_1),\ L_b(\lambda_2) \} = \Big  [r_{ab}(\lambda_1 - \lambda_2),\ L_a(\lambda_1) \ L_b(\lambda_2) \Big ]. \label{rtt}
\ee
In this case the $r$-matrix is the familiar $\mathfrak{sl}_2$ Yangian matrix \cite{yang}:
\be
r(\lambda) = {1\over \lambda}  {\cal P}, \label{rmatrix}
\ee
${\cal P}$ is the permutation operator: ${\cal P} \Big ( \vec{a} \otimes \vec{b} \Big) = \vec{b} \otimes \vec{a}$. The formula (\ref{rtt}) is then realized by the following relations:
\be
\{x_i,\ X_j\} = \delta_{ij} \label{can},
\ee
that is $x,\ X$ are canonical variables.

The discrete model with $N$ sites, and periodic boundary conditions is associated to the transfer matrix defined as \cite{FT, FTS, tak}:
\be
t(\lambda) = Tr_a\ T_a(\lambda) ~~~~\mbox{where} ~~~~~T_a(\lambda) = L_{aN}(\lambda) L_{aN-1}(\lambda) \ldots L_{a1}(\lambda). \label{mono0}
\ee
$T$ is the monodromy matrix also satisfying the quadratic algebraic relation (\ref{rtt}). In the notation $L_{0i}$, the index $a$ denotes the auxiliary space, whereas the index $i$ denotes the $i$th site on the one dimensional lattice derived by (\ref{mono0}). As will be transparent later in the text in the continuum limit the discrete index $i$ will be replaced by the continuum coordinate $x$.

The transfer matrix $t(\lambda)$ as is well known provides all the charges in involution. Indeed, via (\ref{rtt}) one readily shows that
\be
\Big \{ t(\lambda),\ t(\mu)\Big \}=0,
\ee
hence the system is by construction integrable.

To derive the {\it local} integrals of motion one should expand the $\ln t(\lambda)$ in powers of $\lambda$ or ${1 \over \lambda}$. In this case we expand in powers of ${1\over \lambda}$, because the $L$-matrix (\ref{lmatrix}) reduces to the degenerate matrix $D$ at $\lambda \to \infty$.

Let us now expand the monodromy matrix:
\be
T(\lambda \to \infty) & \propto & D_N\ldots D_1  + {1\over \lambda} \sum_{i=1}^N D_N \dots D_{i+1} A_i D_{i-1}
\ldots D_1
\non\\&+& {1\over \lambda^2} \sum_{i>j} D_N \ldots D_{i+1} A_i D_{i-1} \ldots D_{j+1} A_j \ldots D_1
\non\\
&+& {1\over \lambda^3} \sum_{i>j>k} D_N \ldots D_{i+1} A_i D_{i-1} \ldots D_{j+1} A_j
\ldots D_{k+1} A_k \ldots D_1 \non\\ &+& \ldots
\ee
Taking into account the latter expansion and the definition of the transfer matrix we conclude:
\be
\ln t(\lambda \to \infty) \propto {1 \over \lambda} H_1 + {1\over \lambda^2} H_2 + {1\over \lambda^3} H_3 + \ldots,
\ee
where the extracted integrals of motion have the following familiar form (see also e.g. \cite{kundura, doikouit})
\be
&& H_1 = \sum_{i=1}^N {\mathbb N}_i, \non\\
&& H_2 = - \sum_{i=1}^N x_{i+1} X_i - {1\over 2} \sum_{i=1}^N {\mathbb N}_i^2 \non\\
&& H_3 = -\sum_{i=1}^N x_{i+2}X_i + \sum_{i=1}^N ({\mathbb N}_i+{\mathbb N}_{i+1})x_{i+1}X_i
+{1\over 3}\sum_{i=1}^N{\mathbb N}_i^3.
\ee
The latter provide the first integrals of motion (number of particles, momentum and Hamiltonian respectively) of the whole hierarchy for the NLS model. It is clear that the continuum limits of the above quantities provide the corresponding integrals of motion of the continuum NLS model \cite{kundura, doikouit}. The latter expressions are valid in the quantum case as well (see e.g. \cite{kundura, doikouit}).

\subsection{The Lax pair formulation}
Let us now briefly review how the Lax pair associated to each local integral of motion is derived via the $r$-matrix formulation (see also \cite{FT}).
Introduce first the Lax pair ($L,\ {\mathbb A}$) for discrete integrable models, and the
associated discrete auxiliary linear problem (see e.g. \cite{FT})
\be
 && \psi_{j+1} = L_j\ \psi_j \non\\ && \dot{\psi}_j =  {\mathbb A}_j\ \psi_j.
\label{lat}
\ee
From the latter equations one may immediately obtain the discrete zero curvature condition as a compatibility condition:
\be \dot{L}_j=
 {\mathbb A}_{j+1}\ L_j - L_j\  {\mathbb A}_j. \label{zero}
\ee
Recall that the index $j$ denotes the site on an one dimensional lattice, and it will be
replaced in the continuum limit by the continuum coordinate $x$. In the continuum limit as will be clear the equations (\ref{lat}), (\ref{zero}) reduce to the continuum linear auxiliary problem and the continuum zero curvature condition respectively (see section 5).

Let us introduce at this point some useful notation. We define for $i >j$:
\be
T_a(i,j;\lambda)= L_{ai}(\lambda) L_{a i-1}(\lambda) \ldots L_{aj}(\lambda).
\ee
To be able to construct the Lax pair  we should first formulate the following Poisson structure \cite{FT}:
\be
\Big \{ T_a(\lambda),\ L_{bj}(\mu) \Big \} &=& T_a(N,j+1;\lambda)r_{ab}(\lambda-\mu)T_a(j,1;\lambda)L_{bj}(\mu)
\non\\ &-& L_{bj}(\mu) T_a(N,j;\lambda) r_{ab}(\lambda- \mu)T_a(j-1, 1;\lambda).
\ee
It then immediately follows for the generating function of the {\it local} integrals of motion:
\be
\Big \{\ln t(\lambda),\ L_{bj}(\mu)\Big \} &=& t^{-1}(\lambda)\ Tr_a\Big (T_a(N,j+1;\lambda)\ r_{ab}(\lambda-\mu)\ T_a(j,1;\lambda)
\Big )\ L_{bj}(\mu) \non\\ &-&  L_{bj}(\mu)\ t^{-1}(\lambda)\ Tr_a \Big (T_a(N,j;\lambda)\ r_{ab}(\lambda-\mu)\ T_a(j-1,1;\lambda)  \Big). \label{local}
\ee
Recalling the classical equation of motion
\be
\dot{L}_j(\mu) = \Big \{\ln t(\lambda),\ L_j(\mu) \Big \},
\ee
and comparing with expression (\ref{local}) we obtain
\be  {\mathbb A}_j(\lambda, \mu) = t^{-1}(\lambda)\
tr_{a}\ \Big  [ T_a(N,j;\lambda)\ r_{ab}(\lambda -\mu)\ T_a(j-1,1;\lambda) \Big ], \label{laf} \ee
where the relevant classical $r$-matrix is given in (\ref{rmatrix}).

Substituting the $r$-matrix into the latter expression we conclude that
\be
 {\mathbb A}_j(\lambda, \mu) ={t^{-1}(\lambda) \over \lambda -\mu}\ T(j-1, 1;\lambda)\ T(N,j;\lambda).
\ee
Expansion of the latter expression in powers of ${1\over \lambda}$ provides the Lax pairs associated to each one of the
local integrals of motion (see also \cite{avandoikou1}), i.e.:
\be
&& {\mathbb A}_j^{(1)}(\mu) =  \begin{pmatrix}
    1 & 0 \\
    0 & 0
\end{pmatrix}, ~~~~
 {\mathbb A}_j^{(2)}(\mu) =\begin{pmatrix}
   \mu & x_j \\
    -X_{j-1} & 0
\end{pmatrix}, \non\\
&&  {\mathbb A}_j^{(3)}= \begin{pmatrix}
   \mu^2  +x_j X_{j-1} & \mu x_j -x_j {\mathbb N}_j + x_{j+1}\\
    -\mu X_{j-1} + X_{j-1}{\mathbb N}_{j-1}-X_{j-2} & -x_jX_{j-1}
\end{pmatrix}. \label{bulka}
\ee

Both the Lax pair via the zero curvature condition and the Hamiltonian description give rise to the same equations of motion. Consider for instance the equations of motion associated to $H_3$ (and the Lax pair $L,\  {\mathbb A}^{(3)}$). Indeed from
\be
\dot{x}_j = \{H_3,\ x_j\}, ~~~~\dot{X}_j= \{H_3,\ X_j \},
\ee
and via the zero curvature condition for the pair $L,\ {\mathbb A}^{(3)}$ we obtain the following set of difference equations:
\be
\dot{x}_j &=& x_{j+2} - 2x_{j+1} {\mathbb N}_j -  x_{j+1} {\mathbb N}_{j+1 }+ x_j {\mathbb N}_j^2 + x_j^2 X_{j-1} + x_{j+1}\non\\
\dot{X}_j &=& -X_{j-2} +2X_{j-1} {\mathbb N}_j + X_{j-1} {\mathbb N}_{j-1} -X_j {\mathbb N}_j^2 -X_j^2 x_{j+1}- X_{j-1}.
\label{bulke}
\ee
With this we conclude our brief review on the periodic discrete NLS model.

\section{The DNLS model with integrable defect}
We shall henceforth focus on the discrete NLS model in the presence of an integrable defect.
We shall basically extract the local integrals of motion and the corresponding Lax pairs for
the aforementioned model, and shall derive the modified equations of motion due to the presence
of the defect.

Let us first describe the algebraic setting for the defect per se.
Introduce the Lax operator associated to the defect, which is located at a particular site say $n$:
\be
\tilde L_{an} &=& \lambda + \tilde A_{an} \non\\
&=&  \lambda + \begin{pmatrix}
   \alpha_n & \beta_n \\
    \gamma_n & \delta_n
\end{pmatrix} \ ,
\ee
the index $n$ simply denotes the position of the defect on the one dimensional spin chain.
Note that the $\tilde L$ matrix is required to obey the same Poisson bracket structure with the bulk matrices $L$ (\ref{lmatrix}) so that integrability is ensured.
The entries of the above $\tilde L$ matrix  may be parameterized as (see e.g. \cite{ADS}, and references therein)
\be
\alpha_n = -\delta_n = {1\over 2} \cos (2\theta_n), ~~~~~~\beta_n = {1\over 2}\sin(2\theta_n) e^{2i\phi_n},
~~~~\gamma_n = {1\over 2} \sin(2\theta_n) e^{-2i\phi_n} ,
\ee
the fields $\theta_n,\ \phi_n$ may be rewritten in terms of the canonical variables $p_n$ and $q_n$ as
\be
&&\cos(2\theta_n) = p_n, ~~~~~\phi_n =q_n \cr
&&\{q_n,\ p_n\} =i.
\ee
It is then immediately shown via the algebraic relation (\ref{rtt}) that the elements $\alpha_n,\ \beta_n,\ \gamma_n,\ \delta_n$ satisfy the following exchange relations:
\be
&& \{\alpha_n,\ \beta_n\}= \beta_n \non\\
&&\{\alpha_n,\ \gamma_n\} = -\gamma_n \non\\
&& \{\beta_n,\ \gamma_n\} = 2 \alpha_n
\ee
which are the typical $\mathfrak{sl}_2$ exchange relations.

For simplicity, and in order to avoid unwanted boundary effects we shall consider the defect away from the ends of the
one dimensional lattice model.
Inserting the defect at the $n$ site of the one dimensional lattice the corresponding monodromy matrix is expressed as:
\be
T_a(\lambda) = L_{aN}(\lambda) L_{aN-1}(\lambda) \ldots \tilde L_{an}(\lambda) \ldots L_{a1}(\lambda) .
\ee
Note that due to the fact that the $\tilde L$-operator is required to satisfy the same fundamental algebraic relation as the monodromy matrix, the trace of it --the transfer matrix-- provides
a family of Poisson commuting operators. Having the latter expression at our disposal we may now construct the desired physical quantities.

\subsection{Local Integrals of motion}
First we wish to extract the associated local integrals of motion. They are obtained, as in the previous section, from the expansion of  $\ln t(\lambda)$. Let us first present the expansion of the relevant monodromy matrix:
\be
T(\lambda \to \infty) & \propto & D_N\ldots D_1  \non\\
&+& {1\over \lambda} \left (\sum_{n\neq i =1}^N D_N \dots D_{i+1} A_i D_{i-1}
\ldots D_1\ +\  D_N \dots D_{n+1} \tilde A_n D_{n-1}
\ldots D_1 \right )
\non\\
&+& {1\over \lambda^2} \sum_{i>j} D_N \ldots D_{i+1} A_i D_{i-1} \ldots D_{j+1} A_j \ldots D_1
\non\\
&+& {1\over \lambda^2} \sum_{n>j} D_N \ldots D_{n+1} \tilde A_n D_{n-1} \ldots D_{j+1} A_j \ldots D_1
\non\\
&+& {1\over \lambda^2} \sum_{j>n} D_N \ldots D_{j+1} A_j D_{j-1} \ldots D_{n+1} \tilde A_n \ldots D_1
\non\\
&+& \ldots
\ee
The technical details are omitted for brevity, and we directly provide the final expressions:
\be
\log t(\lambda) = {1\over \lambda} {\cal H}_1 +{ 1\over \lambda^2}{\cal H}_2 + { 1\over \lambda^3}{\cal H}_3 + \ldots
\ee
We shall write down here the first three terms of the expansion,
which after some tedious computations are given by (the expressions below hold at the quantum level as well):
\be
{\cal H}_1 &=& \sum_{j\neq n} {\mathbb N}_j + \alpha_n \non\\
{\cal H}_2 &=&   - \sum_{j \neq n, n-1} x_{j+1} X_j-{1\over 2}\sum_{j\neq n}{\mathbb N}_j^2
-x_{n+1}X_{n-1} - \beta_nX_{n-1} + \gamma_n x_{n+1} -{\alpha_n^2 \over 2}\non\\
{\cal H}_3 &=& -\sum_{j \neq n, n\pm 1} x_{j+1} X_{j-1} + \sum_{j \neq n, n-1} ({\mathbb N}_j + {\mathbb N}_{j+1})x_{j+1} X_j +
{1\over 3} \sum_{j\neq n} {\mathbb N}_j^3 + \tilde x_{n, n+1}{\mathbb N}_{n-1}  X_{n-1}\non\\
&+&  \tilde X_{n, n-1} x_{n+1}{\mathbb N}_{n+1} + \alpha_n\tilde x_{n, n+1} X_{n-1}  +\alpha_n \tilde X_{n, n-1} x_{n+1}-\tilde x_{n, n+1} X_{n-2} -x_{n+2} \tilde X_{n, n-1}+{{\alpha_n}^3\over 3} \non\\
\ee
where we define
\be
\tilde x_{n, n+1} &=& x_{n+1} + \beta_n
\non\\
\tilde X_{n, n-1} &=& X_{n-1}- \gamma_n.
\ee
It is clear that as we consider higher orders in the expansion, the terms associated to the defect become less and less local. And although the defect is attached to a particular site $n$, its effect to higher integrals of motion becomes highly non-local. A similar behavior is naturally expected when deriving the relevant Lax pairs as will be transparent in the subsequent section.

\subsection{The associated Lax pair}
In this case one has to distinguish various cases due to the presence of the impurity. More precisely as we consider higher order expressions we need to take into account more and more points around the defect in order to include all the possible interactions. For instance, to derive ${\mathbb A}^{(2)}$ we consider the ``bulk'' points and separately the point $n,\ n+1$. For ${\mathbb A}^{(3)}$ we separately evaluate the operator for the points $n,\ n \pm 1,\ n+2$ an so on. Indeed, the main observation is that the presence of the defect described by the Lax operator $\tilde L_n$ induces non-trivial ``boundary'' type effects onto the neighboring operators ${\mathbb A}_j$ around the defect point. More precisely,  the generic expression for ${\mathbb A}_j$ the sites $n,\ n+1$ or instance are given as:
\be
{\mathbb A}_n(\lambda,\ \mu) &=& {t^{-1}(\lambda) \over \lambda -\mu}\ L_{n-1}(\lambda) \ldots L_1(\lambda)\ L_N(\lambda) \ldots \tilde L_n(\lambda) \non\\
{\mathbb A}_{n+1}(\lambda,\ \mu) &=& {t^{-1}(\lambda) \over \lambda -\mu}\ \tilde L_{n}(\lambda) \ldots L_1(\lambda)\ L_N(\lambda) \ldots  L_{n+1}(\lambda)
\ee
and so on for points around the defect. The non-trivial ``boundary'' effects are due to the fact that the $\tilde L$ operator is located near or on the edges of the sequence of the Lax operators in the latter expressions.

After some quite tedious computations we conclude that: the Lax pair ${\mathbb A}_j^{(1)}$ remains the same as in (\ref{bulka}) for all sites, ${\mathbb A}_j^{(2)}$ for $j \neq n,\ n+1$ is given by expression (\ref{bulka}), whereas
\be
{\mathbb A}_{n}^{(2)} = \begin{pmatrix}
   \mu & \beta_n + x_{n+1} \\
   -X_{n-1} & 0
\end{pmatrix}, ~~~~~{\mathbb A}^{(2)}_{n+1}= \begin{pmatrix}
   \mu & x_{n+1} \\
   \gamma_n -X_{n-1} & 0
\end{pmatrix}
\ee
Also ${\mathbb A}_j^{(3)}$ for $j \neq n,\ n\pm 1,\ n +2 $ is given by (\ref{bulka}) and:
\be
{\mathbb A}_{n-1}^{(3)} &=& \begin{pmatrix}
\mu^2 + x_{n-1} X_{n-2}  &  \mu x_{n-1} +\tilde x_{n, n+1} -{\mathbb N}_{n-1} x_{n-1}\\
-\mu X_{n-2} -X_{n-3} +{\mathbb N}_{n-2} X_{n-2}
 & - X_{n-2}x_{n-1}
\end{pmatrix} \non\\
{\mathbb A}_{n}^{(3)} &=& \begin{pmatrix}
\mu^2 + \tilde x_{n, n+1} X_{n-1} &  \mu \tilde x_{n, n+1} +x_{n+1}
-{\mathbb N}_{n+1} x_{n+1}  +{\mathfrak f} \\
-\mu X_{n-1} -X_{n-2} + {\mathbb N}_{n-1} X_{n-1}  & -\tilde x_{n, n+1} X_{n-1}
\end{pmatrix} \non\\
{\mathbb A}_{n+1}^{(3)} &=& \begin{pmatrix}
\mu^2 + x_{n+1} \tilde X_{n,n-1}   &  \mu x_{n+1} +x_{n+2} -{\mathbb N}_{n+1} x_{n+1}\\
-\mu \tilde X_{n, n-1} -X_{n-1} +{\mathbb N}_{n-1} X_{n-1} +{\mathfrak g}
 &  - \tilde X_{n, n-1}x_{n+1}
\end{pmatrix} \non\\
{\mathbb A}_{n+2}^{(3)} &=& \begin{pmatrix}
\mu^2 + x_{n+2} X_{n+1}  &  \mu x_{n+2} +x_{n+3} -{\mathbb N}_{n+2} x_{n+2} \\
-\mu X_{n+1} -\tilde X_{n, n-1} +{\mathbb N}_{n+1} X_{n+1}
 & - X_{n+1}x_{n+2}
\end{pmatrix} \non\\
\ee
where we define
\be
{\mathfrak f} &=& x_{n+2} - x_{n+1} -\alpha_n (\beta_n +2 x_{n+1}) \non\\
{\mathfrak g} &=& X_{n-1} -X_{n-2} - \alpha_n(\gamma_n - 2 X_{n-1}).
\ee
Having been able to explicitly derive the first integrals of motion as well as the associated Lax pairs we may now
identify the
corresponding difference equations of motion, and check the consistency of the approaches followed. Indeed, both
descriptions, i.e. the Hamiltonian
as well as the zero curvature condition provide as expected the same equations of motion.

Let us now focus on the third charge, and extract the relevant equations of motion. These are given for
$j \neq n,\ n\pm 1,\ n\pm 2$
by equations (\ref{bulke}),
whereas for the points around the impurity we obtain:
\be
\dot{x}_{n-2} &=& \tilde x_{n, n+1} -2 x_{n-1} {\mathbb N}_{n-2} - x_{n-1}{\mathbb N}_{n-1} + x_{n-2}{\mathbb N}_{n-2}^2 +
X_{n-3}x_{n-2}^2 + x_{n-1} \non\\
\dot{X}_{n-2} &=& -X_{n-4} +2X_{n-3}{\mathbb N}_{n-2} +X_{n-3} {\mathbb N}_{n-3} -X_{n-2}{\mathbb N}_{n-2}^2
-x_{n-1}X_{n-2}^2 - X_{n-3}\non\\
\dot{x}_{n-1} &=& x_{n+1} -2\tilde x_{n, n+1}{\mathbb N}_{n-1}  - {\mathbb N}_{n+1}x_{n+1} +x_{n-1}{\mathbb N}_{n-1}^2
+x_{n-1}^2X_{n-2} + \tilde x_{n, n+1} +{\mathfrak f} \non\\
\dot{X}_{n-1} &=& -X_{n-3} +2 X_{n-2}{\mathbb N}_{n-1} +X_{n-2}{\mathbb N}_{n-2} -X_{n-1}{\mathbb N}_{n-1}^2 -
\tilde x_{n, n+1}X_{n-1}^2  - X_{n-2}  \non\\
\dot{x}_{n+1} &=& x_{n+3} -2 x_{n+2}{\mathbb N}_{n+1} - x_{n+2} {\mathbb N}_{n+2} + x_{n+1} {\mathbb N}^2_{n+1}
+x_{n+1}^2 \tilde X_{n, n-1}+  x_{n+2}\non\\
\dot{X}_{n+1} &=& - X_{n-1} +2 \tilde X_{n, n-1} {\mathbb N}_{n+1} -X_{n-1} {\mathbb N}_{n-1}-
X_{n+1} {\mathbb N}^2_{n+1} -x_{n+2} X_{n+1}^2
 -\tilde X_{n, n-1} + {\mathfrak g} \non\\
\dot{x}_{n+2}&=&  x_{n+4} -2x_{n+3}{\mathbb N}_{n+2} -x_{n+3}{\mathbb N}_{n+3} +x_{n+2} {\mathbb N}^2_{n+2} +
x_{n+2}^2 X_{N+1} + x_{n+3}\non\\
\dot{X}_{n+2} &=& - \tilde X_{n, n-1} +2X_{n+1} {\mathbb N}_{n+2} +X_{n+1} {\mathbb N}_{n+1} -X_{n+2}{\mathbb N}_{n+2}^2 -
x_{n+3}X_{n+2}^2 -X_{n+1}. \non\\
\ee
Particular attention is given to the defect point. In this case one has to take into account the defect degrees of freedom and the exchange relations among the elements $\alpha,\ \beta,\ \gamma,\ \delta$ when considering the equations of motion from the Hamiltonian. From the zero curvature condition on the other hand one has to bear in mind that exactly on the defect point the $L$-operator is modified to $\tilde L$, thus the condition may be rewritten as:
\be
\dot{\tilde L}_n(\lambda) = {\mathbb A}_{n+1}(\lambda)\ \tilde L_n(\lambda)- \tilde L_n(\lambda)\ {\mathbb A}_n(\lambda)
\ee
and the entailed equations of motion for the defect point are given as:
\be
\dot{\alpha}_n &=& -\beta_n {\mathbb N}_{n-1} X_{n-1} - \gamma_n x_{n+1} {\mathbb N}_{n+1} +\beta_nX_{n-2} + \gamma_n x_{n+2}
-\alpha_n\beta_nX_{n-1} - \alpha_n \gamma_n x_{n+1} \non\\
\dot{\beta_n} &=& 2\alpha_n x_{n+1} {\mathbb N}_{n+1} -2 \alpha_n x_{n+2} +2 \beta_nx_{n+1}X_{n-1}
+\beta_n^2 X_{n-1}
- \beta_n \gamma_n x_{n+1} +2 \alpha_n^2 x_{n+1}  + \alpha_n^2 \beta_n \non\\
\dot{\gamma}_n &= &2 \alpha_n X_{n-1} {\mathbb N}_{n-1} -2 \alpha_nX_{n-2} - 2 \gamma_n x_{n+1} X_{n-1} -\gamma_n\beta_nX_{n-1}
+2 \alpha_n^2X_{n-1} + \gamma_n^2x_{n+1} - \alpha_n^2\gamma_n. \non
\ee

Taking the continuum limit of the discrete model under study is a significant aspect of the whole process. It is an essential step towards understanding how integrability
can be preserved in the continuum case. There is a discussion on the continuum NLS models in \cite{cozanls},
but there is no convincing argument as far as we can understand on the issue of integrability.
Both descriptions i.e. the Hamiltonian versus the Lax pair formulation are needed in order to
obtain a complete view of the problem at hand.
It is technically more convenient in many cases to use the information from the Lax pair formulation or vise versa,
however in most cases combination of both descriptions helps to completely describe the problem especially when
dealing with the continuum version of a lattice integrable model.

\section{The continuum limit: a first glance}
In order to proceed with the continuum limit of the discrete NLS model let us first
introduce the  spacing parameter $\Delta$ in the $L$-matrix of the discrete NLS
model as well as in the $\tilde L$ matrix of the defect (index free notation):
\be
L(\lambda) =  \begin{pmatrix}
   1+ \Delta \lambda - \Delta^2 x X & \Delta x \\
    -\Delta X  & 1
\end{pmatrix} \
\ee

\be
\tilde L(\lambda) = \Delta \lambda + \begin{pmatrix}
   \alpha & \beta \\
    \gamma & \delta
\end{pmatrix} \
\ee
where  we now define:
\be
\alpha = -\delta= {1\over 2} \cos(2\Delta \theta),  ~~~~\beta ={1\over 2} \sin (2\Delta \theta)e^{2i \phi},
~~~~~ \gamma ={1\over 2} \sin (2\Delta \theta)e^{-2i \phi},
\ee
we also define:
\be
\theta e^{2i\phi} = y, ~~~~\theta e^{-2 i \phi} =Y,
\ee
the latter identifications will be used in the following analysis. Notice that the spectral parameter $\lambda$ is also suitably renormalized to $\Delta \lambda$ in both $L$ and $\tilde L$ matrices in order to formulate a sensible continuum limit process compatible also with the continuum linear algebra (see also \cite{ADS}). Moreover, such a renormalization is necessary if we wish the whole process to be compatible with the so called ``power counting'' argument introduced in \cite{ADS}.

Having introduced the appropriate spacing parameter in the $L$-operators above we may now consider the continuum limit of the integrals of motion of discrete NLS models and the associated Lax pairs. Before obtaining the continuum limit let us first introduce the following notation. In particular, we set:
\be
&&  x_j\ \to\ x^-(x), ~~~~~X_j\ \to\ X^-(x), ~~~~~1 \leq j \leq n-1, ~~~~~x\in (-\infty,\ x_0) \cr
&&  x_j \to x^+(x), ~~~~~ X_j  \to X^+(x), ~~~~~~n+1 \leq j \leq N, ~~~~~x \in (x_0,\ \infty).
\ee
where $x_0$ is the defect position in the continuum theory.
Note also that in order to perform the continuum limit we bear in mind that:
\be
&& \Delta\ \sum_{j=1}^{n-1} f_j \ \to\ \int_{-\infty}^{x_0^-}dx\ f^-(x) \non\\
&& \Delta\ \sum_{j=n+1}^{N} f_j \ \to\ \int_{x_0^+}^{\infty} dx\ f^+(x).
\ee

The continuum limit of the first integral of motion is then given as:
\be
{\cal H}^{(1)} = - \int_{-\infty}^{x_0^-} dx\ x^{-}(x) X^-(x) - \int^{\infty}_{x_0^+} dx\ x^{+}(x) X^+(x).
\ee
Notice that in the first integral we considered terms proportional to $\Delta$, whereas in the second integral the first non trivial contribution to the continuum limit is of order $\Delta^2$. The respective continuum quantity reads then as
\be
{\cal H}^{(2)} &=&  -\int_{-\infty}^{x_0^-} dx\ x^{-'}(x) X^-(x) - \int^{\infty}_{x_0^+}dx\ x^{+'}(x) X^+(x) \non\\ & +& x^-(x_0) X^-(x_0)- x^+(x_0) X^-(x_0) + x^+(x_0)Y(x_0) -y(x_0) X^-(x_0) +{1\over 2} y(x_0)  \non\\
\ee
the prime denotes derivative with respect to $x$.

Note that in the continuum limit\footnote{Notice that the $\tilde L$ matrix in the continuum limit may be expressed as:
\be
\tilde L \sim \sigma_3+\Delta\ \tilde U
\ee
$\sigma_3$ the familiar Pauli matrix.
We could have chosen instead $\bar L = \sigma_3 \tilde L$, which also satisfies the quadratic relation (\ref{rtt}), and has the expected continuum behavior (\ref{contb}). Such a choice would slightly modify the defect terms in the local integrals of motion. Note that such modifications can be suitably implemented in the continuum monodromy matrix, but we shall discuss this matter in detail elsewhere.}:
\be
L(\lambda) \sim {\mathbb I} + \Delta\ {\mathbb U}(\lambda) \label{contb}
\ee
also the continuum zero curvature condition with Lax pair ${\mathbb U},\ {\mathbb V}$ takes the from:
\be
\dot{\mathbb U} - {\mathbb V}' +\Big [{\mathbb U},\ {\mathbb V} \Big ] =0.
\ee
The Lax pair associated to the first integral in quite trivial and coincides with the one in (\ref{bulka}).
The Lax pair associated to the second integral of motion is given by the following expressions:
\be
&&{\mathbb V}^{(2)}(\mu,\ x) = \begin{pmatrix}
   \mu & x^-(x) \\
    -X^-(x) & 0
\end{pmatrix} \  ~~~x \in (-\infty,\ x_0^-], \cr
&&{\mathbb V}^{(2)}(\mu,\ x) = \begin{pmatrix}
   \mu & x^+(x) \\
    -X^+(x) & 0
\end{pmatrix} \  ~~~x \in (x^+_0,\ \infty) \cr
&&{\mathbb V}^{(2)}(\mu,\ x_0) =\begin{pmatrix}
   \mu & x^+(x_0)+ y(x_0) \\
    -X^-(x_0) & 0
\end{pmatrix} \ ,\cr
&& {\mathbb V}^{(2)}(\mu,\ x_0^+)=\begin{pmatrix}
   \mu & x^+(x_0) \\
    Y(x_0) -X^-(x_0) & 0
\end{pmatrix} \ .
\ee
Due to continuity requirements at the points $x_0^+,\ x_0^-$ (see also a similar argument in \cite{avandoikou1}, we end up with the following sewing conditions associated to the defect point:
\be
&& y(x_0) = x^-(x_0) - x^+(x_0), \cr && Y(x_0) = X^-(x_0)- X^+(x_0). \label{sew}
\ee
Notice that the continuity argument may be successfully applied to the points around the defect, however as expected  a discontinuity (jump) is observed exactly on the defect point. It should be emphasized that the $L$ operator is altered at $x_0$ (i.e. $L\ \to\ \tilde L$), leading to modification or discontinuity in the zero curvature condition at $x_0$, which accordingly lead to adjustments in the induced equations of motion.

It is also quite straightforward to show that if the sewing conditions (\ref{sew}) are valid then
\be
\{ {\cal H}_1,\ {\cal H}_2 \} =0,\label{com2}
\ee
which is a first good indication of the preservation of the integrability in the continuum case as well. However, this is somehow ``on shell'' information, given that one requires (\ref{sew}) in order to prove the Poisson commutativity (\ref{com2}). Moreover, the constraints (\ref{sew}) provide a first hint on the existence of an underlying non-ultra local algebra (see e.g. \cite{fredmai}) associated to the defect point. And although in the discrete case one deals with an ultra local algebra there is an indication that in the continuum limit one has to consider a generalized non-ultra local algebra in order to efficiently describe the point like defect at $x_0$.

Here, we only provide a first glimpse on the continuum limit of the discrete NLS models. To obtain the continuum counterparts of the higher integrals of motion and the associated Lax pairs requires subtle manipulations. Such an explicit construction is beyond the intended scope of the present investigation, however we shall analyze this intriguing issue in full detail in future works \cite{avandoikouprep}.

\section{Discussion}
Let us briefly summarize the main findings of the present study. The main aim of this work was the investigation of the discrete NLS model in the present of an ultra local integrable defect. Based on purely algebraic considerations we were able to extract the first charges in involution for the discrete model. The model is by construction integrable given that the bulk $L$ matrices as well as the Lax matrix associated to the defect are required to obey the same Poisson bracket structure.

Then, again by exploiting the underlying algebra, we extracted the associated Lax pairs. Particular attention was given to the construction of the Lax pairs around the defect point. It turned out that the behavior of more and more points around the defect is affected as we move to higher order expressions. Having this information at our disposal we were able to derive the sets of the difference equations of motion. Finally, we provided a first insight of the continuum behavior of the system by considering the continuum limit of the first two integrals of motion. This led to certain sewing or compatibility conditions that ensure Poisson commutativity of the first two integrals of motion at the continuum case as well.

It is worth noting that at this stage it is difficult to conclude whether or not our results correspond to the results of e.g. \cite{cozanls} or if any comparison whatsoever can be made, given that the issue of integrability is still open in \cite{cozanls}. The systematic continuum limit of discrete models in the presence of integrable defects in the spirit of \cite{ADS} is our next target \cite{avandoikouprep}, and it will provide a deeper understanding on the connection with earlier works. In fact, this systematic process will lead to the derivation of the continuum limit of higher integrals of motion, such as ${\cal H}^{(3)}$ and the corresponding Lax pairs, as well as the associated constraints (sewing conditions). Note that the entailed higher constraints will involve as expected spatial derivatives. Moreover, Poisson commutativity of all the entailed charges needs to be explicitly checked so that we can claim that integrability holds. Compatibility of the higher sewing conditions should be also explicitly checked. These are highly non-trivial technical points, and will be presented in full detail in future investigations.

Similar ideas may be put forward in the case of other well known prototype models such as the Heisenberg model (see e.g. \cite{weston}) aiming also at the investigation of the corresponding continuum theories.
In any case, a detailed analysis on continuum integrable models in the presence of defects turns out to be a fundamental issue, which will be addressed in forthcoming publications.

\paragraph{Acknowledgements\\}
I am indebted to J. Avan for illuminating discussions, useful suggestions, and ongoing collaboration on this subject.


\begin{thebibliography}{99}

\bibitem{delmusi}
G. Delfino, G. Mussardo and P. Simonetti, Phys. Lett. B328 (1994) 123, {\tt hep-th/9403049};\\
G. Delfino, G. Mussardo and P. Simonetti, Nucl. Phys. B432 (1994) 518, {\tt hep-th/9409076}.

\bibitem{fus} E. Corrigan and C. Zambon, J. Phys. A: Math. Theor. 43 (2010) 345201, {\tt arXiv:1006.0939 [hep-th]}.

\bibitem{konle}
R. Konik and A. LeClair, Nucl. Phys B538 (1999) 587; {\tt hep-th/9793985}.

\bibitem{BCZ1}
P. Bowcock, E. Corrigan and C. Zambon, JHEP 08(2005) 023, {\tt hep-th/0506169}.

\bibitem{nemes}
F. Nemes, Semiclassical analysis of defect sine-Gordon theory, Int. J. Mod. Phys. A 25
(2010) 4493; arXiv:0909.3268 [hep-th].

\bibitem{BCZ2}
E. Corrigan and C. Zambon, J. Phys. A
42 (2009) 304008, {\tt arXiv:0902.1307 [hep-th]};\\
P. Bowcock, E. Corrigan and C. Zambon, JHEP 01
(2004) 056, {\tt hep-th/0401020}

\bibitem{BCZ3}
E. Corrigan and C. Zambon,
JHEP 07 (2007) 001, {\tt arXiv:0705.1066 [hep-th]};\\
E. Corrigan and C. Zambon, J. Phys. A 42 (2009) 475203;
{\tt arXiv:0908.3126 [hep-th]}.

\bibitem{cozanls}
E. Corrigan and C. Zambon, Nonlinearity 19 (2006) 1447, {\tt nlin/0512038}.

\bibitem{haku}
I. Habibullin and A. Kundu, Nucl. Phys. B 795 (2008) 549, {\tt arXiv:0709.4611 [hep-th]}.

\bibitem{caudr} V. Caudrelier, IJGMMP vol.5, No. 7 (2008) 1085.

\bibitem{weston}
R. Weston, {\it An Algebraic Setting for Defects in the XXZ and Sine-Gordon Models}, {\tt arXiv:0065369 [math-ph]}.

\bibitem{kundura}
A. Kundu and O. Ragnisco, J. Phys. A27 (1994) 6335, {\tt hep-th/9401066}.

\bibitem{FT}
L.D. Faddeev and L.A. Takhtakajan, {\it Hamiltonian Methods in the Theory of Solitons},
(1987) Springer-Verlag.

\bibitem{yang}
C.N. Yang, Phys. Rev. Lett. {\bf 19} (1967) 1312.

\bibitem{FTS}
L. Faddeev, E. Sklyanin and L. Takhtajan, Theor. Math. Phys. 40 (1980) 688;\\
N. Yu. Reshethikhin, L. Takhtajan and L.D. Faddeev, Len. Math. J. 1 (1990) 193.

\bibitem{tak}
L.A. Takhtajan, {\it Quamtum Groups}, Introduction to Quantum Groups and
Intergable Massive models of Quantum Field Theory, eds, M.-L. Ge and B.-H. Zhao, Nankai
Lectures on Mathematical Physics, World Scientific, 1990, p.p. 69.

\bibitem{doikouit}
A. Doikou, D. Fioravanti and F. Ravanini, Nucl. Phys. B790 (2008) 465, {\tt arXiv:0706.15.15 [hep-th]}.

\bibitem{avandoikou1}
J. Avan and A. Doikou, Nucl. Phys. B812 (2009) 481, {\tt arXiv:0809.2734 [hep-th]}.

\bibitem{ADS}
J. Avan, A. Doikou and K. Sfetsos, Nucl. Phys. B840  (2010) 469, {\tt arXiv:1005.4605 [hep-th]}.

\bibitem{fredmai}
L. Freidel and J.M. Maillet, Phys. Lett. B262 (1991) 278.

\bibitem{avandoikouprep}
J. Avan and A. Doikou, work in progress.

\end{thebibliography}
\end{document}